\documentstyle[12pt,twoside]{article}
\def\greaterthansquiggle{\raise.3ex\hbox{$>$\kern-.75em\lower1ex\hbox{$\sim$}}}
\def\lessthansquiggle{\raise.3ex\hbox{$<$\kern-.75em\lower1ex\hbox{$\sim$}}}
\newcommand{\beq}{\begin{equation}}
\newcommand{\eeq}{\end{equation}}
\newcommand{\beqa}{\begin{eqnarray}}
\newcommand{\eeqa}{\end{eqnarray}}
\newcommand{\beqan}{\begin{eqnarray*}}
\newcommand{\eeqan}{\end{eqnarray*}}
\newcommand{\ba}{\begin{array}}
\newcommand{\ea}{\end{array}}

\newcommand{\A}{{\cal A}}

\newcommand{\C}{{\cal C}}

\def\nz{\ifmmode {I\hskip -3pt N} \else {\hbox {$I\hskip -3pt N$}}\fi}
\def\zz{\ifmmode {Z\hskip -4.8pt Z} \else
       {\hbox {$Z\hskip -4.8pt Z$}}\fi}
\def\qz{\ifmmode {Q\hskip -5.0pt\vrule height6.0pt depth 0pt
       \hskip 6pt} \else {\hbox
       {$Q\hskip -5.0pt\vrule height6.0pt depth 0pt\hskip 6pt$}}\fi}
\def\rz{\ifmmode {I\hskip -3pt R} \else {\hbox {$I\hskip -3pt R$}}\fi}
\def\cz{\ifmmode {C\hskip -4.8pt\vrule height5.8pt\hskip 6.3pt} \else
       {\hbox {$C\hskip -4.8pt\vrule height5.8pt\hskip 6.3pt$}}\fi}

\def\au{{\setbox0=\hbox{\lower1.36775ex%
\hbox{''}\kern-.05em}\dp0=.36775ex\hskip0pt\box0}}
\def\ao{{}\kern-.10em\hbox{``}}

\normalsize
\begin{document}
\bibliographystyle{plain}

\begin{titlepage}
\begin{flushright}
\today
\end{flushright}
\vspace*{2.2cm}
\begin{center}
{\Large \bf Local Normality for KMS States with Galilei invariant interaction}\\[30pt]

Heide Narnhofer  $^\ast $\\ [10pt] {\small\it}
Fakult\"at f\"ur Physik \\ Universit\"at Wien\\

\vfill \vspace{0.4cm}

\begin{abstract}For Galilei invariant Fermi systems interacting by a pair potential with a cut off for particles with highly different velocity the time evolution corresponds to an automorphism. We prove that all states satisfying the KMS condition with respect to this interaction are locally normal to the representation in Fock space. Removing this cut off limit states remain locally normal provided the interaction is repulsive or of positive type. The modular automorphism on the von Neumann algebra of the limit state coincides with the time evolution.

\\
\hspace{1.9cm}

\end{abstract}
\end{center}

\vfill {\footnotesize}

$^\ast$ {E--mail address: heide.narnhofer@
univie.ac.at}
\end{titlepage}
\section{Introduction}
In \cite{HHW} the physical system is described as a $C^*$-algebra with quasilocal structure on which the time evolution acts as a continuous automorphism group. Then the equilibrium states are those that satisfy the KMS-condition. Special examples of such $C^*$-algebras are the $C^*$-algebras over an infinite lattice where the local algebras are finite dimensional algebras over a lattice point. In these models the existence of the dynamics can be proven under appropriate conditions on the interaction, as well as the existence of KMS-states together with relevant results on their structure and properties \cite{BR}, \cite{AF}. These models can be generalized to the $C^*$-algebra built by fermionic creation and annihilation operators over lattices in a way, that the essential results for the lattice algebras remain valid \cite{AM}.

However in physics we are interested in continuous systems. Here it is our hope that again most of the considerations and results for lattice algebras keep their relevance. We need control over the time evolution and we want control on equilibrium states telling us that locally it is sufficient to concentrate on few particles and on few modes.

Already for a finite number of particles the appropriate choice of a $C^*$-algebra is not evident. In  \cite{FV} it is shown that for a finite number of particles the $C^*$-algebra built by Weyl-operators is invariant under the free time evolution, but the evolution including a potential is only defined for the $W^*$-closure. In \cite{BG} and concentrating on the nonrelativistic situation in \cite{DB} it is suggested to choose another algebra, i.e. the resolvent algebra, for which the evolution including a potential acts as automorphism.  \cite{GS} however suggested to start from the very beginning with a $C^*$-algebra that is the norm closure of an increasing net of $W^*$-algebras, locally given as the $W^*$-algebra over the Fock-space. Instead of considering all possible states it should be sufficient to concentrate only on those states whose GNS representations are locally normal. The task is then to prove that for such a system the dynamics exists and that the set of permitted states is invariant under this dynamics.
It is easy to construct dynamics with point interaction between the particles such that the dynamics is not stable in the sense that the hamiltonian for a fixed particle number is not bounded from below by the particle number. As a consequence no ground state exists, particles can accumulate decreasing the potential energy while increasing the kinetic energy, so that locally we do not stay in Fock-space. Computer simulations support that this can happen even in finite time.

In order to inherit as much wisdom as possible from the lattice algebras we considered in \cite{NT} the $C^*$-algebra built by fermionic creation and annihilation operators over the continuum, but did not consider point interaction between the particles but an interaction built by products of creation and annihilation operators over coherent vectors in a way that the interaction vanishes when the particles are far apart or when they have highly different velocities.  Letting the cut-off in velocity space tend to infinity the interaction converges to the familiar point interaction in Fock-space with slightly smeared potential on the level of the time derivative, but this does not suffice to pass to the time evolution as an automorphism group. Keeping the cut-off in these models this allows to use estimates similar as in classical phase space so that perturbation theory works similar as for the interaction in the lattice models. Therefore the time evolution exists as automorphism group over the $C^*$-algebra, the system is stable in the sense that in Fock-space the hamiltonian is bounded from below by the number operator, and it allows the construction of KMS-states for all temperatures and all chemical potentials.

In this note we will show that these KMS-states are locally normal to states in Fock space in support of the starting point in \cite{GS}. The argument is based on a characterization of KMS-states by the time derivative. With this characterization we can show that the expectation value of particles with high velocity vanishes sufficiently fast to guarantee local normality. These KMS-states depend on the cut-off parameter. Considering the limit of these KMS-states when the cut-off parameter tends to zero, we have to control for which potentials the limit state remains locally normal. We concentrate on repulsive interaction, whereas the same argument fails for attractive interactions, supporting the expectation that stability is necessary for local normality. Using the resulting convergence of the time derivative on the local level we can apply the results of \cite{DB}, that this derivative defines an automorphismgroup on an appropriately constructed algebra. Since in locally normal states the weak closures of the algebras coincide this also shows that the choice of the $C^*$-algebra is not relevant in the sense of \cite{GS} for equilibrium states, provided the time evolution is defined by the time derivative given on the local level.

\section{The basic inequality}
In \cite{BR}, based on results in \cite{AS}, \cite{S} and \cite{Ro}, KMS states are characterized by the following
\paragraph{Theorem 1:}
Let $(\A, \tau _t )$ be a $C^*$ dynamical system, $\delta $ the infinitesimal generator of $\tau _t $, and $\omega $ a state over $\A.$ $\omega $ is a $(\beta, \tau _t )$-KMS state, iff
\beq  -i\beta \omega (A^*\delta(A))=i\beta \omega (\delta (A^*)A)\geq \omega (A^*A) \log \Big(\frac{\omega (A^*A)}{\omega (AA^*)}\Big)\eeq
with the understanding that \beq u\log\frac{u}{v}=0 , u=0,v\geq0;\quad u\log\frac{u}{v}=\infty,  u>0,v=0.\eeq
Whereas in \cite{AS} and \cite{S} this theorem was used to relate the KMS condition to thermodynamical functions on a local basis, \cite{Ro} applied it to get information on equilibrium states. They were interested in lattice systems. But the result holds also for Fermi systems where the time evolution can be extended by the gauge automorphism for the field algebra, i. e. for including the action on odd operators. Especially, for quasifree Fermi systems it can be used to obtain the KMS-states. We will generalize the idea to control also interacting systems. We are guided by the fact, that for matrix algebras the time derivative permits eigenoperators, and for these eigenoperators the inequality (1) reduces to an equality and defines the state. For infinite systems such eigenoperators do not exist. However, if it is possible to find suitable approximations for them, (1) offers the possibility to approximate also their expectation-value in KMS states. In the following we will concentrate on $\beta =1$. Generalisations to general $\beta $ are possible by scaling arguments with the exception $\beta=\infty$, which corresponds to the groundstate. For $\beta =0$ the inequality defines
 the tracial state, which exists for the $C^*$-algebra but is not locally normal.
\paragraph{Theorem 2:}
Let $\omega $ be a KMS-state for $\beta =1$ with respect to the time automorphism group $\tau _t$ and corresponding time derivative $\delta.$ Assume there exists a sequence of operators $A_{\lambda },||A_{\lambda }||=1$, and $B_{\lambda }$ such that $-i\delta A_{\lambda }==-\lambda A_{\lambda }+B_{\lambda, }, \quad 0<\lambda  \rightarrow \infty,$ where $||B_{\lambda }||\leq h(\lambda )$. Assume that
\beq  h(\lambda )\leq c .\eeq
Then
\beq \lim _{\lambda \rightarrow \infty }\omega (A^*_{\lambda }A_{\lambda })\lambda^2/(h^2+1)=0  \quad  \forall\epsilon >0.\eeq
Proof: With the ansatz $\omega (A^*_{\lambda }A_{\lambda })=1/y$ and using Schwartz inequality the inequality (1) implies the inequality
\beq f(\lambda)=\lambda - hy^{1/2}(\lambda )- \log y(\lambda ) \leq 0.\eeq
 Assume $y(\lambda )$ is smaller than $\lambda ^2/(h^2 +1) $. Then $f(\lambda )$ becomes positive for large $\lambda$ and the inequality is violated.
By a similar estimate we can improve the estimate on the decrease property of $1/y$: for $h(\lambda )=e^{-d\lambda }, 0<d<1$,  $\omega (A^*_{\lambda }A_{\lambda })$ decreases like $e^{-d\lambda }.$  If $d>1$, $h(\lambda )$ is not the dominating factor any more and $\omega (A^*_{\lambda }A_{\lambda })$ decreases like $e^{-\lambda }.$ If $h(\lambda )$ is just bounded we keep with (4).

The condition on $B_{\lambda }$ allows variations. For example we can also use
\beq A^*_{\lambda }B_{\lambda }+B_{\lambda }^*A_{\lambda }<cA^*_{\lambda }A_{\lambda }+D_{\lambda },\quad ||D_{\lambda }||<h(\lambda ) ,\eeq
so that in the estimate $\lambda $ can be replaced by $\lambda -c.$

Another alternative splits \beq i(A^*_{\lambda }\delta A_{\lambda }-\delta A^*_{\lambda } A_{\lambda })=2A^*_{\lambda }V_{\lambda }A_{\lambda }-A^*_{\lambda }A_{\lambda }V_{\lambda }-V_{\lambda }A^*_{\lambda }A_{\lambda } -2\lambda A^*_{\lambda }A_{\lambda }\eeq
for an appropriate $V_{\lambda }, B_{\lambda }=i[V_{\lambda}, A_{\lambda }]$ and estimates \beq |\omega (A_{\lambda }^*A_{\lambda }V_{\lambda })|=|\langle \Omega|A_{\lambda }V_{\lambda }e^{-H}A^*_{\lambda }|\Omega \rangle |\leq \eeq
$$ \langle \Omega|A_{\lambda }e^{-H}A^*_{\lambda }|\Omega \rangle ||e^{H/2}V_{\lambda }e^{-H/2}||= \omega (A^*_{\lambda }A_{\lambda })||e^{H/2}V_{\lambda }e^{-H/2}||$$
where $e^{iHt}$ implements the KMS-automorphism and satisfies $H|\Omega \rangle =0.$ Similarly
\beq \omega (V_{\lambda }A^*_{\lambda }A_{\lambda })\leq \omega (A^*_{\lambda }A_{\lambda })||e^{-H/2}V_{\lambda }e^{H/2}|| \eeq
\beq \omega (A^*_{\lambda }V_{\lambda }A_{\lambda })\leq \omega (A^*_{\lambda }A_{\lambda })||V_{\lambda }|| .\eeq
If $V_{\lambda }$ is an analytic element of the algebra of observables this again implies just a shift in the estimate and we get exponential decrease.
\paragraph{Example :}
In order to see the power of the theorem we apply it to fermionic annihilation operators $a(f),f\in L^2(\mathrm{R})$ and consider the free time evolution combined with the gauge automorphism in order to describe KMS-states with given chemical potential $\mu.$ Take $a(f_{p_0,c}
)$ to be the annihilation operator concentrated around the momentum $p_0$ and let $c$ characterize the concentration such that \beq i\delta a(f_{p_0,c}))=(p_0^2+\mu)a(f_{p_0,c})+B_{c,p_0} \eeq
$$\quad B_{c,p_0}= a(g_{p_0}) \quad g_{p_0}(p)=(p_0^2-p^2)f_{p_0,c}(p), \quad   \lim _{c\rightarrow \infty }||B_{c,p_0}||\rightarrow 0  $$
where we can make the convergence of $B$ as fast as we want. Then with $$\omega
\Big(a^*(f_{p_0,c})a(f_{p_0 ,c})\Big)=w(p_0,c)$$

\beq p_0^2+\mu +b(c)\geq \log(\frac{1-w}{w}),\eeq

Applying  theorem 1 to the creation operator and using (1) and the anticommutation relation between creation and annihilation operators gives
\beq p_0^2+\mu -b(c)\leq \log(\frac{1-w}{w}).\eeq
Together with $\lim _{c \rightarrow \infty }||B_{c,p_0}||=\lim _{c \rightarrow \infty } b(c)=0$ we get the well known formula
\beq \lim _{c\rightarrow \infty } w(p_0,c)=\frac{1}{1+e^{p_0^2+\mu }}.\eeq

\section{A Toy Model}
We consider a simplified version of the model presented in \cite{B}. The total algebra is a quasilocal algebra over a one-dimensional lattice system. The local algebra $\A _n$ at the point $n$ is a $C^*$ algebra imbedded in an infinite Hilbert space and built by the operators $|k\rangle \langle l|_n$ where the vectors $|k\rangle  $ form an orthonormal basis in this Hilbert space . We assume that the interaction is restricted to nearest neighbors and we make the ansatz
\beq H_N=\sum _{n=-N}^N \Big(...\otimes 1_{n-1}\sum _k k|k\rangle \langle k|_n\otimes1_{n+1}.. +\sum _{klrm}..\otimes 1_{n-1} \otimes h_{klrm}|k\rangle \langle r|_n\otimes |l\rangle \langle m|_{n+1} \otimes 1_{n+2}..+\eeq
$$\sum _{klrm}..\otimes 1_{n-2}\otimes h_{klrm}|l\rangle \langle m|_{n-1} \otimes |k\rangle \langle r|_n\otimes  1_{n+1}...\Big)$$
where we assume periodic boundary conditions. Time evolution is given for the limit $N\rightarrow \infty.$ Other boundary conditions are also possible as long as they define locally the same time derivative in the limit $N\rightarrow \infty $. These boundary conditions  may become relevant for the limit of the KMS- states. But every such limit state has to satisfy (1) with respect to
 the time derivative acting on local operators. We are interested in its action on  local matrix units where
\beq \delta (..1_{-1} \otimes |k\rangle \langle 0|_0 \otimes 1_1...)=...1_{-1}\otimes k|k\rangle \langle 0|_0\otimes 1_1.. +..\sum _{rlm}..h_{rlkm}|l\rangle \langle m|_{-1} \otimes |r\rangle \langle 0|_0 \otimes 1_1...\eeq
$$-\sum _{lsm}..h_{0lsm}... 1_{-1}\otimes |k\rangle \langle s|_0 )\otimes (|l \rangle \langle m|_1 ..$$
The representation of a state $\omega $ is locally normal, if the algebra $\omega $ restricted to $\A_n$ is still represented by a density matrix expressed by one-dimensional matrix units $|k\rangle \langle l|_n$ in the infinite Hilbert space. This holds, if $\omega (|k\rangle \langle k|_n)$ vanishes sufficiently fast by a similar but simpler proof as given for lemma 3 and lemma 4 below.
In order to apply theorem 2  we collect the various contributions in $-i\omega (..\otimes |k\rangle\langle0|\otimes..\delta (..\otimes |0\rangle \langle
k| \otimes..))$. They  have the form \beq \label{1} -..\otimes k|k\rangle \langle k|_0 \otimes ..,\eeq
\beq \label{2} h_{0l0m}..\otimes |k \rangle \langle k|_0 \otimes |l \rangle \langle m|_1...,\eeq
\beq \label{3} h_{klrm}..\otimes |k \rangle \langle r|_0 \otimes |l\rangle \langle m|_1.. ,\eeq
\beq \label{4} h_{0lm0}...\otimes |l\rangle \langle m|_{-1}\otimes |k \rangle \langle k|_0 \otimes ..\eeq
\beq \label{5} h_{klrm}...\otimes |l\rangle \langle m|_{-1}\otimes |k \rangle \langle r|_0 \otimes ..\eeq
  The contributions from  (18) and (20) in (1) are dominated by (17) if $\sum _{lm}|h_{0l0m}|<c$ and in the sense of (6) act as a shift $\lambda (k)=k-c$ in theorem2. (19) and (21) correspond to $D_{\lambda }$ in (6) and we  have to demand that
\beq \sum _{lrm}|h_{klrm}|<ck^{\alpha }, \quad \alpha \leq 0,\eeq
where $\alpha $ influences the decay properties.

 If $\omega $ is a limit state of $\omega _N$ where $\omega _N$ is a KMS state with respect to $H_N$ then (1) is satisfied for local operators. So the conditions in Theorem 2 are met also for the weak*-limit $N\rightarrow \infty$ and we can conclude that $$\lim_{K_0 \rightarrow \infty }\sum _{k>K_0} \omega (..\otimes |k\rangle \langle k|_0\otimes ..)=0.$$ The restriction of $\omega $  to $\A_n$ is then given by a density operator and the convergence properties can further be used to give an upper bound on the local entropy. From the subadditivity of the entropy it follows that also the entropy for $\A_{[-l ,l]}$ is finite and the state is normal for every local algebra.

We did not control whether $H_N$ defines in the limit $N\rightarrow \infty $ an automorphism on the chosen $C^*$ algebra. But the limit state satisfies the condition on the derivative for a densely defined set of operators and according to Theorem 1 we interpret it as an equilibrium state.

\section{Interacting Fermi Systems}
For an interacting system in the groundstate, increasing the number of particles can lead to divergencies if the interaction is not stable. The kinetic energy may increase as the potential energy decreases. This does not happen when the potential energy is bounded from below by the kinetic energy together with the number operator. Then we can add to the hamiltonian  the number operator with an appropriate chemical potential so that this operator is bounded from below in Fock-space and permits to consider the grand canonical ensemble. We concentrate on fermionic systems and consider the $C^*-$algebra built by creation and annihilation operators., so that in all estimates we can use, that they are norm bounded.  In \cite{NT} we interpreted the assumption of stability as a restriction, that potential energy cannot be shifted without limitation to the kinetic energy and expressed this limitation by cutting off the interaction for particles with relatively high velocities. In addition we tried to be as close as possible to reality by keeping Galilei invariance. This could be realized by choosing as the hamiltonian
\beq
\label {FG}H= \frac{1}{2m}\int dx^{\nu}
\nabla  a^*(x)\nabla a(x)+\int d^{\nu}pd^{\nu}p'd^{\nu}qd^{\nu}q' a^*_{pq}a^*_{p'q'}w(p-p')v(q-q')a_{p'q'} a_{pq}=K+V\eeq
in $\nu $ dimensions. Here $a_{pq}=a(W(p,q)f)$ is an annihilation operator smeared with an $f$ that is translated by the one-particle Weyl operators $W(p,q)=e^{i(qP+pX)}.$ As concrete example we take as in \cite{NT} $f$ as a Gauss function so that with appropriate normalization in $\nu $ dimensions  with the notation
$$[a^*(x),a(y)]_+=\delta ^{\nu }(x-y)$$
$$a(f(x))=a(f)=\int d^{\nu} x f(x)a(x)$$
$$a_{pq}=\pi^{-\frac{\nu}{4}}\int d^{\nu }xe^{-\frac{(q-x)^2}{2}+ipx}a(x)=a(|p,q\rangle )$$
where $|p,q\rangle $ are coherent states.

In the following we will omit $\nu $ if it is of no relevance. This hamiltonian defines an automorphism group that resembles the time evolution of the Fermi system on the lattice: we can expand in $t$ and can control the convergence \cite{NT}. The advantage of the above model is the fact that  the time evolution is related to the space translation by Galilei invariance so that it can inherit weak asymptotical abelianess \cite{Nwa}. Galilei invariance connects space, boost, gauge and time transformations
 \beq\sigma _xa(f(y))=a(f(x+y))\quad \gamma _b a(f(y))=a(e^{iby}f(y))\quad
\nu _{\alpha }a(f)=e^{i\alpha }a(f) \quad
 \tau _t,\eeq
such that
$$ \sigma _x \circ \nu _{\alpha } =\nu _{\alpha } \circ \sigma _x, \quad \gamma _b \circ \nu _{\alpha } =\nu _{\alpha} \circ \gamma _b, \quad \gamma _b \circ \sigma _x=\sigma _x \circ \gamma _b \circ \nu _{-bx}$$
$$ \tau _t \circ \nu _{\alpha }=\nu _{\alpha } \circ \tau _t, \quad \tau _t \circ \sigma _x =\sigma _x \circ \tau _t \quad \tau _t \circ \gamma _b =\gamma _b \circ \tau _t \circ \sigma _{bt} \circ \nu _{-b^2t/2}.$$
It replaces Poincare-invariance in relativistic theories and is therefore a necessary demand, if we want to learn from the model what might be essential for local normality also in this context.

If we integrate only over a finite region in $p,p', q,q',$ the interaction becomes a norm bounded perturbation of the free evolution. Starting with a KMS-state for the free evolution we begin with this bounded perturbation staying in the representation and then let the integration regions tend to infinity. As the dynamics is converging in norm, by the  Banach-Alaoglu theorem   weak$^*$ limit points of these states exist and satisfy the KMS condition with respect to $\tau $ (including the chemical potential) \cite{NT} (corollary 4.10).   Now we can state
\paragraph{Theorem 3:}
Translation invariant KMS-states for the dynamics $\tau _t\circ \nu _{\mu t}$, given by the hamiltonian (24) and chemical potential corresponding to adding $\mu N$ to the hamiltonian, are locally normal to the representation in Fock- space.

To prove this statement we first control that because of Theorem 2 the two-point function decreases sufficiently rapidly in $p$, and then we argue that this decrease guarantees local normality.

As for noninteracting systems, we start with $A=a(f_{p_0,c})$, where again $f$ are Gauss functions  with concentration depending on $c$. Then, we only have to show that
\paragraph{Lemma 1:}
The norm of the commutator with the interaction is bounded
\beq||[V,a(f_{p_0,c})]|| \leq g\eeq

Proof: This estimate is the starting point in \cite{NT} for the proof of the existence of the time automorphism. We collect the arguments. Notice that due to the Galilei invariance of the interaction the bound has to be invariant under the action of a Galilei transformation on $a(f(p_0,c)$ , therefore uniform in $p_0.$
The relevant factor in the estimate is the anticommutator between two operators $[a(f_{p_0,c})),a^*_{q'p'}]_+$. With our choice for $f$ as a Gauss function  the anticommutator gives
\beq c(\nu)\int dre^{-(r-p')^2+iq'r-c(r-p_0)^2}=g(c)e^{-\frac{c}{c+1}(p_0-p')^2+iq'(p'+p_0)}=g(p_0,q',p',c).\eeq
where $g(c)$ is dimension dependent and $g(p_0, q',p', c)$ is depending on the indicated parameters, bounded and vanishing integrably for $p'-p_0.$ The commutator with the interaction can be expressed as integral over norm bounded creation and annihilation operators
\beq [V,a(f_{p_0,c})]=\int dpdqa^*_{pq}a_{pq}a(g_{p_0,p,q,})\eeq
where $a(g_{p_0,p,q,})$ is the result of the integral of the remaining $$\int dp'dq'v(q-q')w(p-p')a_{p'q'}g(p_0,q',p',c).$$ The norm of (28), using $||a_{pq}||=1$, can be estimated by
\beq \int dpdq ||a(g_{p_0,p,q,})||=\int dpdq||g_{p_0,p,q,}||_2.\eeq
The function on $r$ $g_{p_0,p,q}(r)$ in (27) is given by $$ \int dp'dq'v(q-q')w(p-p')g(p_0,q',p',c)e^{-(r-p')^2+iq'r}.$$
$||a(g_{p_0,p,q})||=||g_{p_0,p,q}(r)||_2$ is finite and integrably decreasing in $p,q.$ which gives the desired bound on $g$ in(26).

\paragraph{Lemma 2:}
In a translationally invariant KMS state with respect to the dynamics (24) and the chemical potential $\mu $ the two point function (compare (12))
\beq  \bar{w}(p,q)=\omega \Big(a^*_{p,q}a_{p,q}\Big)\eeq
is an $L_1$ function in $p$.

Proof: Due to Lemma 1 $\omega \Big(a^*_{p,q}a_{p,q}\Big)$ satisfies the demands of theorem2. According to (4) $\bar{w}(p,q)$ decays at least as $p^{-4}$ and is therefore integrable in dimensions $\nu \leq 3.$

We have to transfer this estimate to local regions. We concentrate on three dimensions and take as local region a cubus of unit length. Generalizations follow by scaling.

\paragraph {Lemma 3:}
Consider the restriction of the above KMS state on the algebra $\A_{[0,1]^3}.$ With
\beq \omega (a^*(f)a(f))=\langle f|\rho |f\rangle  \quad f\in L^2([0,2 \pi]^3) \eeq
$\rho $ is a density operator.

Proof: $\Pi _{j=1}^3 \cos n_jx_j$ form an orthonormal basis $f_{n_j}$ in $ L^2([0,1]^3)$.  With writing $\Pi _{j=1}^3 \cos (n_jx_j)=|n\rangle $ and using the expression of the identity by coherent states $$\int dpdq|p,q\rangle \langle p,q|=1$$
$$\int _0^{2\pi } d^3 x \Pi _{j=1}^3 \cos n_jx_j f_{p,q}(x)=f(n,p,q)=\langle n|p,q\rangle  $$
$f(n,p,q)$ is a function integrable in $q$ and in $p$ and concentrated at $n-p$. Therefore  we obtain
\beq \omega (a^*(\Pi _{j=1}^3 \cos (n_jx_j))a(\Pi _{k=1}^3 \cos (n_kx_k))=\mathrm{Tr} \rho  |n\rangle \langle n|=\eeq
$$=\int dpdp'dqdq'\mathrm{Tr} \rho |p,q\rangle \langle p,q|n\rangle \langle n|p',q'\rangle \langle p',q'|=$$
$$=\int dpdp'dqdq' \langle p,q|n\rangle \langle n|p',q'\rangle \omega (a^*_{pq}a_{p'q'})$$
which is summable in $n_1,n_2,n_3$ due to the decrease properties of $\langle n|p',q'\rangle $ and $\omega (a^*_{pq}a_{p'q'}).$

Finally this guarantees that the number operator exists and therefore according to \cite{ADR} the state is locally normal and corresponds to a density matrix $\rho _{\Lambda }$ for a local region. We give an alternative proof, that allows also to estimate the entropy $S(\rho _{\Lambda })=-\mathrm{Tr} \rho _{\Lambda }log \rho _{\Lambda }$:
\paragraph{Lemma 4:}
Let $\omega $ be a state over the algebra built by creation and annihilation operators $a^*(f),a(f)$. Assume that its two-point function can be expressed by a positive compact operator
\beq\omega \Big(a^*(f)a(f)\Big)=\langle f|\rho |f\rangle ,\eeq
where $\rho \leq 1,$ $\rho =\sum \rho _n|f_n\rangle \langle f_n|$, $ \rho _n<cn^{-(1+\epsilon)}$ for some basis $\{f_n \}$. Then the GNS representation of $\omega $ is normal to the Fock representation and its entropy can be estimated from above by the entropy corresponding to the two-point function.

Proof: Take $n=1,2...N.$ Let $R_N$ be the algebra built by $a(f_n),1\leq n\leq N.$ Consider the quasifree state $\tilde{\omega }_N$ over this algebra given by the above two-point function. This state factorizes into states over the algebras $\A_n $ built by $a(f_n).$ For this algebra $P_n=a^*(f_n)a(f_n)$ and $Q_n=1-a^*(f_n)a(f_n)$ are minimal projectors. The products of these projectors form a maximal abelian subalgebra $R_{N0}$ of $R_N$ with minimal projectors $P_{\Pi ,K}$, where $K$ is the number of $P_n$ and $N-K$ is the number of $Q_n$ contributing to $P_{\Pi, K}$ and $\Pi$ indicates the necessary permutations between the sets. The  entropy of $\tilde{\omega }_N$ is given by the sum of the entropy of the factors
$$S(\tilde{\omega }_N)=-\sum _j(\rho _n \log \rho _n+(1-\rho _n)\log (1-\rho _n)).$$
The state $\omega _N$ defined as the restriction of $\omega $ to $R_N$ corresponds to a density matrix $\rho _N $. The maximal abelian algebra consisting of the minimal projections $P_{\Pi, K}$  can be used to define the density matrix $\tilde{\rho }_N=\sum _{\Pi,K} P_{\Pi,K} \rho _N P_{\Pi, K}$, which is the density matrix corresponding to $\tilde{\omega }_N$. Now we use that reduced to our algebra of minimal projectors $\rho _n$ and $\tilde{\rho }_N$ coincide. Then the relative entropy \cite{OP} gives
\beq 0\leq S(\tilde{\rho }_N|\rho _N)=\mathrm{Tr} (\rho _N(\log \rho _N -\log \tilde{\rho }_N)=\mathrm{Tr} \rho _N\log \rho _N -\mathrm{Tr} \tilde{\rho }_N \log \tilde{\rho }_N .\eeq

Therefore $ S(\omega _N)\leq S(\tilde{\omega }_N)$ and the entropy of a state is dominated by the entropy of the restriction of the state to a maximal abelian algebra. Finally
$$ S (\omega )_{\A}=\lim _{N \rightarrow \infty }S(\omega _N)\leq \lim _{N\rightarrow \infty }S(\tilde{\omega }_N)=S(\tilde{\omega })_{\A}.$$
Since according to Lemma 2 and Lemma 3 the two point functions of the KMS-state define a quasifree state that satisfies the condition of Lemma 4 it follows that the restriction of the KMS-state $\omega $ to a local algebra permits a density operator with finite entropy in Fock-space and its representation is therefore locally normal.

\section{Removing the momentum cutoff}
We introduced smeared annihilation operators $a_{pq}$ and the momentum-cutoff, i.e. we added $w_{\gamma}(p-p')=e^{-\gamma (p-p')^2}$ in order to be close to lattice models and to use similar methods to construct the time evolution.
If we want to study an interaction
\beq V=\int dx dx' v(x-y)a^*(x)a^*(y)a(y)a(x)\eeq
we can consider $\lim_{\gamma \rightarrow 0} w_{\gamma }(p-p')=1$ such that in $\nu $ dimension
\beq \int dp a^*_{pq}a_{pq}=c(\nu )e^{-2(q-x)^2}a^*(x)a(x)\eeq
and we obtain a point interaction that is the smeared version of the initial interaction
\beq \tilde{v}(x-x')= \int dqdq'e^{-2(q-x)^2 -2(q'-x)^2 }v(q-q').\eeq
In this sense we define a temperature state corresponding to $\tilde{v}(x-x')$ as a weak*-limit point according to the Banach-Alaoglu theorem of the temperature states corresponding to (24) for the limit $\lim_{\gamma \rightarrow 0} w_{\gamma }(p-p')=1$. If this limit state is locally normal then locally the interaction converges to the point interaction. If it is not locally normal, this state is unphysical, a point interaction is not defined and we have to conclude, that for the corresponding interaction no temperature state exists. As already stated in the introduction we expect that only for one sign of the interaction we can obtain local normality. We consider in the equilibrium state $\omega _{\gamma }$  the term in (1) resulting from the interaction
\beq\int dpdp'dqdq' \omega_{\gamma} (a^*_{p_0,0}a^* _{p,q'}a_{p,q'} a_{p',q} \langle p_0,0|p',q \rangle +a^*_{p',q}\langle p',q|p_0,0\rangle a^* _{p,q'}a_{p,q'} a_{p_0,0}) v(q-q')w_{\gamma }(p-p') .\eeq
It depends linearly on $V$ and can diverge with $\gamma \rightarrow 0.$  If it adds to the contribution of the kinetic energy $p_0^2$, so if the interaction is essentially repulsive , it improves the estimate, otherwise it destroys the estimate, we loose the control and we cannot expect local normality.
  We  concentrate on \beq \int dpdq'a^* _{p,q'}a_{p,q'}v(q-q')w_{\gamma }(p-p')=A_{\gamma }(q,p')\eeq
  so that for $\gamma \rightarrow 0$ according to (35) it converges to \beq c(\nu)e^{-2(q-x)^2}a^*(x)a(x)v(q-x) \eeq
  which for $v\geq 0$ is a positive operator. We rewrite with doing the appropriate integration
  \beq \int dp'dqdq'\omega (a^*(x)(a^*_{p_0,0}a_{p',q'}\langle p',q'|p_o,0\rangle +h.c.)a(x))v(q-q')e^{-(q-x)^2}dx=\eeq
$$=\int dq'dx\omega (a^*(x)a^*_{p_0,0}a(g(p_0,q'))+h.c.)a(x))V_f (x-q')$$
with $V_f$ the appropriately smeared version of $v$ depending on $f$ and approaching $v$ with increasing concentration of $f$ around $p_o$. This corresponds to the expectation of an operator of the form
\beq a^*(h)a(h)+\alpha (a^*(h)a(g)+a^*(g)a(h)) \eeq
The operator that might violate positivity reduces to \beq \alpha (a^*(h+g)a(h+g)-a^*(h-g)a(h-g))\eeq
and will contribute only negligible if the two operators have essentially the same expectation value. In order to control this contribution we write (39) as
\beq \int dxdydz \omega (a^*(x)a^*(y)a(z)a(x))V_f(x-y)V_f(x-z)e^{ip_0(y-z)}f_{p_0}(y)f_{p_0}(z)\eeq
With \beq V_f(x-y)+V_f(x-z)=\sqrt{V_f(x-y)V_v(x-z)}-(\sqrt{V_f(x-y)}-\sqrt{V_f(x-z)})^2 \eeq
the last term tends to $0$ to the amount how well $e^{ip_0(y-z)}f_{p_0}(y)f_{p_0}(z)$ converges to $\delta (y-z)$ for $p_0 \rightarrow \infty$ with respect to $V_f$ that itself converges to $V.$
Therefore we remain with
\beq \int dxdy\omega (a^*(x)a^*(y)a(y)a(x))V_f(x-y)f^2_{p_0}(y)\eeq

which for repulsive potentials or for potentials of positive type is bounded from below and therefore improves the estimate and supports local normality. For attractive potentials it destroys the estimate and we have to expect that $\omega _{\gamma}$ does not converge to a locally normal state.

\paragraph{Theorem 4:}
For repulsive potentials or potentials of positive type weak*-limit points of KMS-states $\omega _{\gamma }$ for $\gamma \rightarrow 0 $ are locally normal. Equilibrium states of systems with repulsive point interaction are locally normal.

\section {The limit states as KMS-states}
In Theorem(1) it was essential that the derivative was the generator of a time automorphism since the proof was based on the existence of analytic elements. Convergence of the states together with local normality guarantees the existence of the derivative given by the point interaction in correspondence to a quadratic form. This is not sufficient to pass to a unique hamiltonian nor that this hamiltonian implements an automorphism that coincides with the modular automorphism so that the time correlations show the usual analyticity properties.  This can be seen in simple counter examples on the basis of quasifree evolution \cite{N}. However local normality allows that we can consider an other net of increasing local algebras. We have to demand that these local algebras are weakly dense in local Fockspace and that for this net the time evolution corresponding to the hamiltonian with point interaction is given as an automorphism. Such a local net is offered in \cite{DB}.

In fact \cite{DB} concentrates on a net for bosons, because he is interested to control unbounded operators by replacing them by their resolvents.  Passing to fermions we have only to replace in his construction the projection onto the symmetric local Fock space in \cite{DB} by a projection onto the antisymmetric one. Estimates remain unchanged or become even simpler. The so constructed algebra $\bar{\A}_{\Lambda }$ is built by compact operators in the local Fock space which in locally normal representations are weakly dense. The total algebra $\bar{\A}$ is now constructed as increasing net of compact operators with a closure given by seminorms, that reflect the imbedding structure but again allow to argue on the basis of Banach Alaoglu when considering sequences of states. For this net it is shown that the time evolution is given as automorphism group. Therefore it is possible to reduce the interaction to local regions, consider the corresponding KMS- state and taking the appropriate weak*-limit point. They are known to exist, but we are interested in controlling their properties.

 Due to local normality the  limit state that we obtained for the $C^*-$ algebra of creation and annihilation operators $\C$
defines density matrices for the local Fock space and therefore defines a state also for the local  $\bar{\A_{\Lambda }}$ and  by taking the limit $\Lambda \rightarrow \mathrm{R}^3$ for the total algebra $\bar{\A }.$ Now the time derivative is the same for $\bar{\A_{\Lambda }}$ as for the local algebra built by creation and annihilation operators and all conditions of Theorem 1 are met. The state is a KMS state for $\bar{\A }$, and also $\pi _{\omega }\bar{\A }''=\pi _{\omega }\C''$ .
 On the level of the $C^*-$ algebra built by creation and annihilation operators point interactions do not allow to construct a time automorphism group. However it is possible, on the basis of (23) and (38) and taking the limit for repulsive interactions, to construct states, that are locally normal, for which the time derivative is defined on a dense set of operators in such a way  that the derivative corresponds to a time automorphism group on the weak closure of the local net of von Neumann algebras, namely the KMS-automorphism corresponding to the state. Time correlations satisfy the analyticity properties of a thermal state. Therefore the limit state can be considered as a KMS-state with respect to the time evolution given by a repulsive point interaction and satisfies all consequences that are implemented by the KMS-condition. Also the description in \cite{GS} is met and we can consider a net of local von Neumann algebras on which the equilibrium state and the time evolution is defined.

\section{Conclusion}
Based on an interaction that is not strictly local but cut off for particles with highly different velocities it is possible to calculate the time evolution for infinite systems of fermions by expansion in the coupling constant of the interaction. An essential consequence is the stability of the interaction, independent of the sign of the interaction. The same idea can be applied for estimating the influence of the interaction on the two point function in temperature states. Again norm estimates are sufficient to guarantee that the free evolution dominates the behaviour for high velocities in all equilibrium states. Therefore the particle density remains finite and temperature states are locally normal with respect to the ground state. The essential ingredient in the estimates is the existence of  an operator, for which the time evolution acts nearly multiplicative. In the present situation this holds for the annihilation operators with high velocity. For such operators the characterization of KMS-states  \cite{AS},\cite{S} gives sufficient bounds on the decrease properties of the expectation values of these operators. The inequalities have  the advantage that the temperature enters just as a multiplication parameter and not on a perturbative level, so that we could use them to prove local normality for all temperatures except $\beta =0$.

Removing the cut off of the interaction in the momentum space destroys norm estimates. We expect that equilibrium states are only locally normal for the appropriate sign of the interaction, for which the system remains stable  such that a ground state with respect to a  dynamics that includes the chemical potential is locally normal to the Fock representation . In agreement with this expectation we show  that at least for repulsive interactions  when removing the momentum cut-off the states remain locally normal. The arguments used for the proof fail for unstable interactions. Finally we use the results in \cite{DB} to deduce that in these limit states the time correlations satisfy the usual analyticity properties of modular time automorphisms.

The analysis is based on norm estimates and therefore not applicable to Bosons. In a toy model we represented how the method works for matrix units respectively for compact operators. Such compact operators are offered in \cite{DB} also for Bosons. There remains the problem whether it is possible to find compact operators that satisfy the conditions of Theorem 2. Note that in \cite{DB} the algebra is gauge invariant whereas for free Bosons we have to use the chemical potential in an appropriate limit to control that the particle density in the Bose condensate remains finite. Additional ideas for understanding the effect of Bose condensation are therefore needed.

\bibliographystyle{plain}

\end{document}